\documentclass[amsmath,amssymb,aps,prd,showpacs,showkeys,data,nofootinbib]{revtex4}
\usepackage[utf8]{inputenc}
\usepackage{graphicx}
\usepackage{amsfonts}
\usepackage{amssymb}
\usepackage{amsmath}
\usepackage{float}
\usepackage{dcolumn}
\usepackage{bm}
\usepackage{latexsym,color}
\begin{document}
\title{Fundamental Frequencies in the Schwarzschild Spacetime
}
\author{
Boshkayev K.A., Muccino M., Rueda J.A., Zhumakhanova G.D.
}

\affiliation{
Dipartamento di Fisico and ICRA Sapienza Universita di Roma Piazzale Aldo Moro 5, I-00185 Rome, Italy ICRANet Piazz della Repubblica 10, I-65122 Pescara, Italy IETP, Faculty of Physics and Technology, Al-Farabi Kazakh National University, Al-Farabi av. 71, 050040 Almaty Kazakhstan
}

\begin{abstract}
We consider the Keplerian, radial and vertical fundamental frequencies in the Schwarzschild spacetime to study the so-called kilohertz quasi-periodic oscillations from low-mass X-ray binary systems. We show that, within the Relativistic Precession Model, the interpretation of observed kilohertz quasi-periodic oscillations in terms of the fundamental frequencies of test particles in the Schwarzschild spacetime, allows one to infer the total mass $M$ of the central object, the internal $R_{in}$ and external $R_{ex}$ radii of accretion disks, and innermost stable circular orbits $r_{ISCO}$ for test particles in a low-mass X-ray binary system. By constructing the relation between the upper and lower frequencies and exploiting the quasi-periodic oscillation data of the Z and Atoll sources we perform the nonlinear model fit procedure and estimate the mass of the central object. Knowing the value of the mass we calculate the internal $R_{in}$ and external $R_{ex}$ radii of accretion disks and show that they are larger than $r_{ISCO}$, what was expected.. 

\end{abstract}

\maketitle

\section{Introduction}

According to the models studying and explaining the physical properties and origin of the quasi-periodic oscillations (QPOs) discovered in low-mass X-ray binary (LMXRB) systems, the QPOs data could provide independent information about the mass of the central compact object (a white dwarf, neutron star and black hole), hence give some clues about the structure of accretion disks \cite{1, 2}. Moreover it is believed that the QPO data might allow one to test the effects of General Relativity (GR) in the strong field regime \cite{3}.  

There are several models attempting to explain the QPOs in the literature \cite{4,5,6,7,8,9,10,11,12}. In all of these models the key point is the assumption that the QPOs originate in the orbits around the compact object onto which the matter accretes. In this paper we will refer to the model that assumes that the QPOs are caused by the fundamental (epicyclic) frequencies associated with the orbital motion of the matter in the accretion disc, such as the Relativistic Precession Model (RPM) \cite{13}. 

The RMP has been put forward in a series of papers \cite{14,15,16}. It explains the QPOs as a direct manifestation of models of relativistic epicyclic motion of test particles at various radii $r$ in theinner parts of the accretion disk. The model identifies the lower (periastron precession $f_{per}$) and upper (Keplerian $f_{K}$) frequencies. In the past years, the RPM has been considered among the candidates to explain the twin-peak QPOs in several LMXBs, and the related constraints on the sources have been discussed \cite{17,18}. 

For test particles in circular orbits, the epicyclic frequency is the frequency of radial and polar motions due to a small perturbation in the orbit. For example, consider the motion of a particle in the field of a compact star in a circular orbit. If the motion of this particle is slightly perturbed in the radial direction, its orbit will still be circular with small oscillations. The epicyclic frequency is the frequency of these oscillations \cite{19}.

The properties of congruencies of nearly circular geodesic orbits in a static and spherically symmetric spacetime such as the Schwarzschild metric are studied because of their fundamental role in the theory of accretion disks around compact objects with strong gravity. The radial epicyclic frequency and vertical epicyclic frequency are the most important characteristics of these orbits.

Analytic formulas for the frequencies in the Schwarzschild, Kerr and Hartle-Thorne metrics have been published many times by several authors \cite{19,20} and are well known. In this paper we consider the frequencies for the Schwarzschild metric in the equatorial plane, and we perform an analysis of a mass estimate carried out in Ref \cite{15}. We show that good fits can be reached leading to an inference of of the model parameters; the mass,external and internal radii of the accretion disk. From this we can obtain the innermost stable circular orbit radius since it depends only on the mass in the case of the Schwarzschild  metric.

\section{The Schwarzschild Spacetime}

In Einstein's theory of general relativity, the Schwarzschild metric is the solution to the gravitational field equations that describes the exterior gravitational field of mass M, based on the assumptions that the electric charge, angular momentum, and other parameters of the source are zero. The Schwarzschild solution was the first exact solution of the Einstein field equations other than the trivial flat space solution.

In ($ct$, $r$, $\theta$, $\phi$) coordinates with signature (+, -, -, -) the Schwarzschild line element has the form

\begin{equation}
\label{1-1}
ds^2=\left(1-\frac{r_g}{r}\right)c^2dt^2-\left(1-\frac{r_g}{r}\right)^{-1}dr^2-r^2\left(d\theta^2+\sin^2\theta d\phi^2\right)
\end{equation} 

where $r_g=2GM/c^2$ is the Schwarzschild radius, $M$ is the mass of the source, $c$ is speed of light in vacuum and $G$ is the gravitational constant \cite{21,22}.

\textit{The fundamental (epicyclic) frequencies}.
For the Schwarzschild spacetime in the equatorial plane $\theta=\pi/2$ the Keplerian frequency for test particles in circular orbits is defined as

\begin{equation} \label{1-2}
f_K=\frac{1}{2\pi}\sqrt{\frac{GM}{r^3}}
\end{equation}

The radial and vertical frequencies of oscillations are given by 

\begin{equation} \label{1-3}
f_r=\frac{1}{2\pi}\sqrt{\frac{GM}{r^3}\left(1-\frac{6GM}{c^2R})\right)}
\end{equation}

\begin{equation} \label{1-4}
f_{\theta}=\frac{1}{2\pi}\sqrt{\frac{GM}{r^3}}
\end{equation}

As one can see the Keplerian and vertical frequencies are equivalent to each other \cite{19}. However in more complex spacetimes they differ. In the RPM the upper frequency is defined as the Keplerian frequency $f_U=f_K$ and the lower frequency is defined as the periastron frequency i.e. $f_L=f_{per}=f_K-f_r$. To perform the non-linear model fit it is convenient to express the lower frequency as a function of the upper frequency $f_L(f_U)$

\begin{equation} \label{1-5}
f_L=f_U \left(1-\sqrt{1-\frac{6(2\pi f_U GM)}{c^2}^{2/3}}\right)
\end{equation}

One may call this expression as the fitting function and here only the mass of the central compact object is a free parameter. The lower and upper frequencies are provided through observations. Knowing the QPO data with their error bars and performing a non-linear model fit procedure one can easily infer the mass with its error bars.

The non-linear model fit routine is based on the least-squares techniques. When there are explicit errors in the data, we use the inverse of the squared errors as fitting weights and form the corresponding chi-squared function of the model parameters. The best-fit parameters are those which minimize this function. In general for a non-linear fit there may be various local minima and therefore an iterative Levenberg-Marquardt algorithm, which starts from an initial guess of the parameter vector, is needed.
\textit{The innermost stable circular orbits.}
Let us consider the motion of a particle in a centrally symmetric gravitational field of a central massive body. As in every centrally symmetric gravitational field, the motion occurs in a single “plane” passing through the origin; we choose the equatorial plane with $\theta=\pi /2$, for the sake of simplicity.

The radius of the innermost stable circular orbits is defined as

\begin{equation}
\label{1-6}
r_{ISCO}=3r_g=\frac{6GM}{c^2}
\end{equation}

$r_{ISCO}$ is considered to be the closest radius around a black hole where one can still have stable circular orbits for test particles \cite{23}. Hence we use it here as a reference to make sure that the inner radius of the accretion disk is larger than $r_{ISCO}$.

\section{Results and discussions}

Formula (5) is the fitting function that we exploit to infer masses for Z (Cir X-1, GX 5-1, GX 17+2) and Atoll (4U1608-52, 4U1728-34) sources. The QPOs data with error bars have been taken from \cite{18,24,25,26,27} and references therein.

\begin{figure}[H]
	\centering
	\begin{tabular}{lr}
		\includegraphics[width=9 cm, height=6 cm, clip]{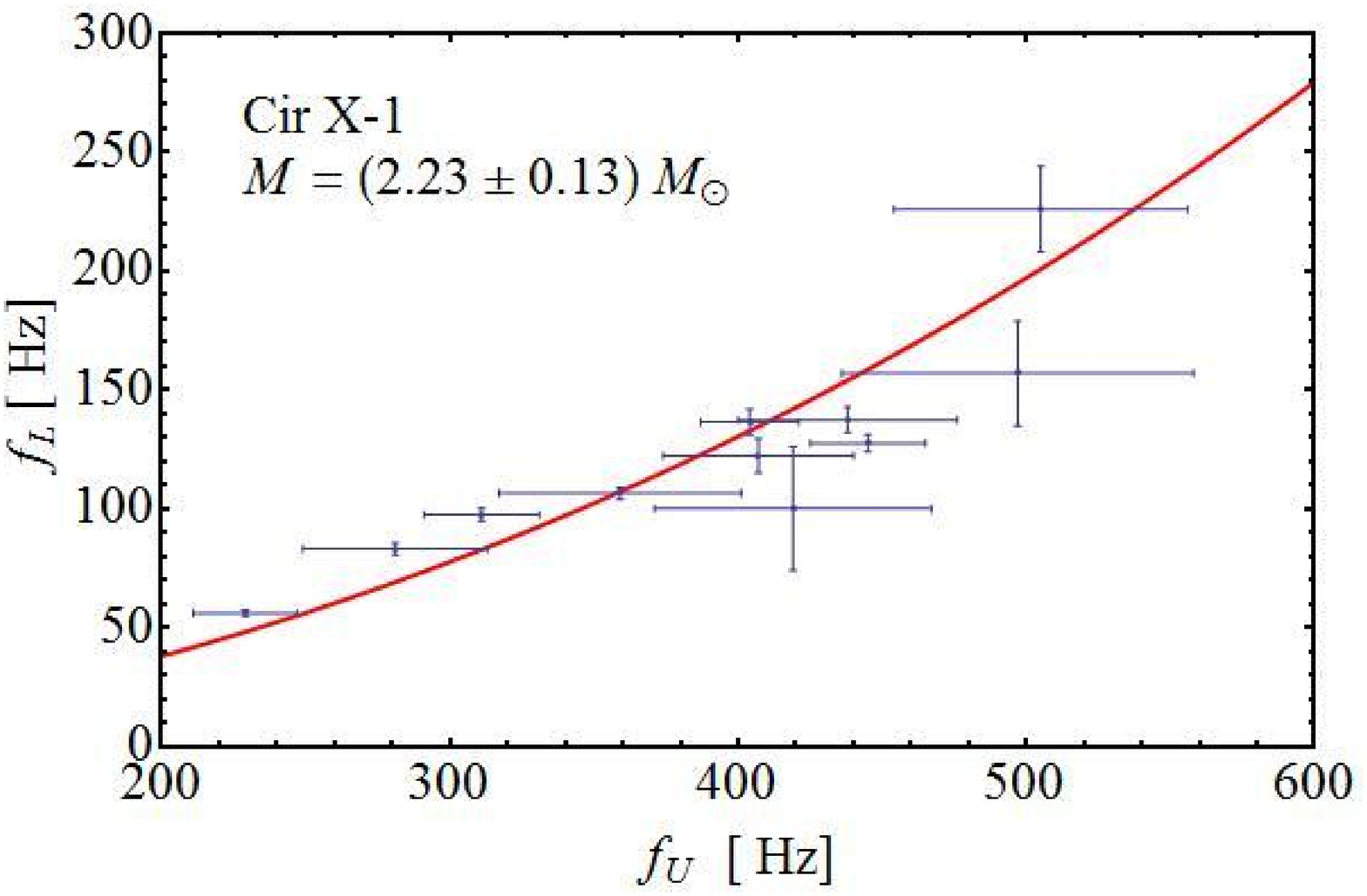}
		&
		\includegraphics[width=8.5 cm, height=6 cm, clip]{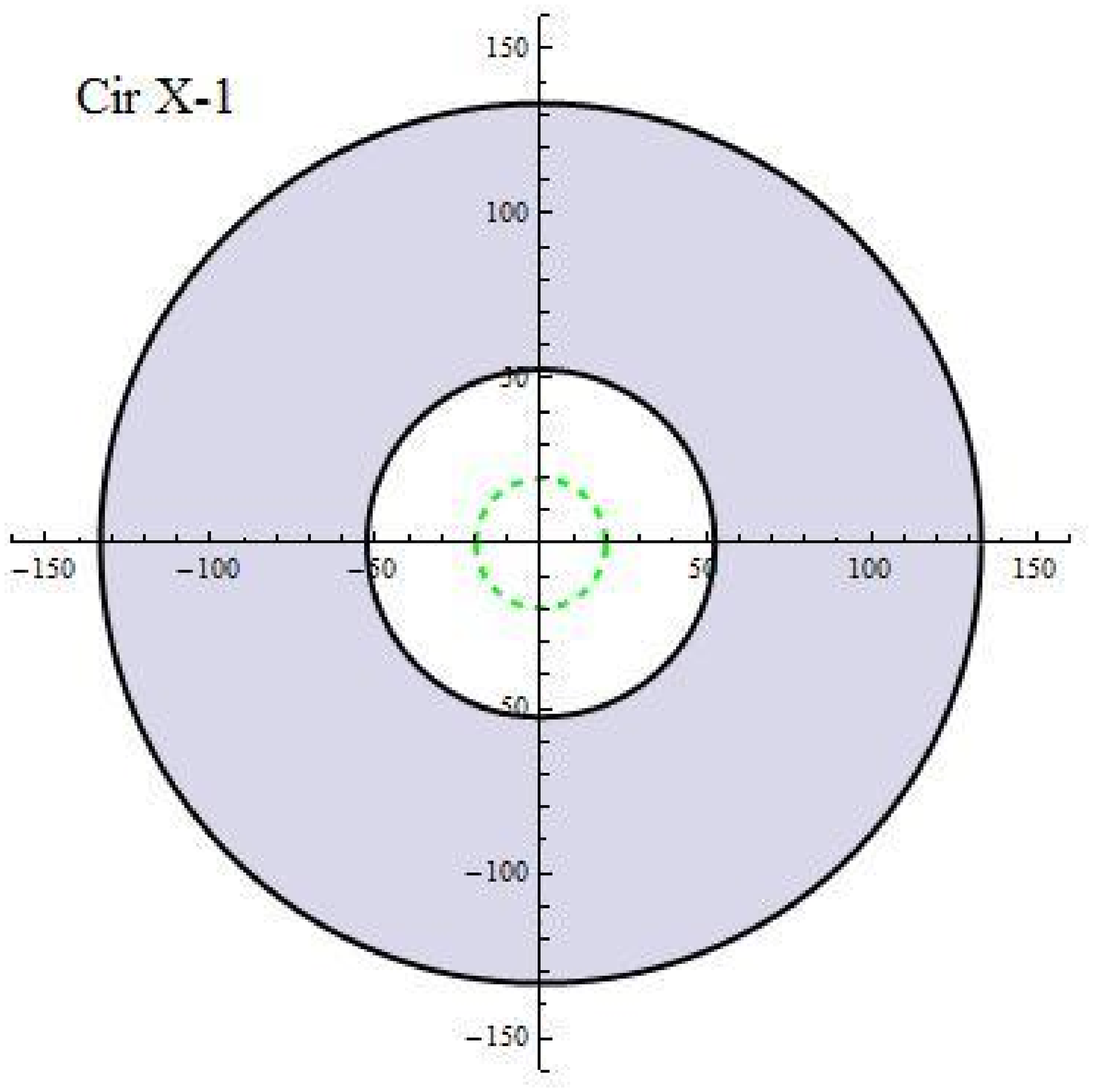}
	\end{tabular}
	\caption{Left panel: the lower frequency ($f_L$) is shown as a function of the upper frequency ($f_U$) for low mass X-ray binary system Cir X-1. Right panel: schematic illustration of the accretion disk around central compact object of Cir X-1. Green dashed circle indicates the radius (in kilometers) of the innermost stable circular orbit ($r_{ISCO}=6GM/c^2$), the inner black circle shows the inner radius, whereas the outer black circle represents the outer edge of the accretion disk.}
	
\end{figure}
\begin{figure}[H]
	\centering
	\begin{tabular}{lr}
		\includegraphics[width=9 cm, height=6 cm, clip]{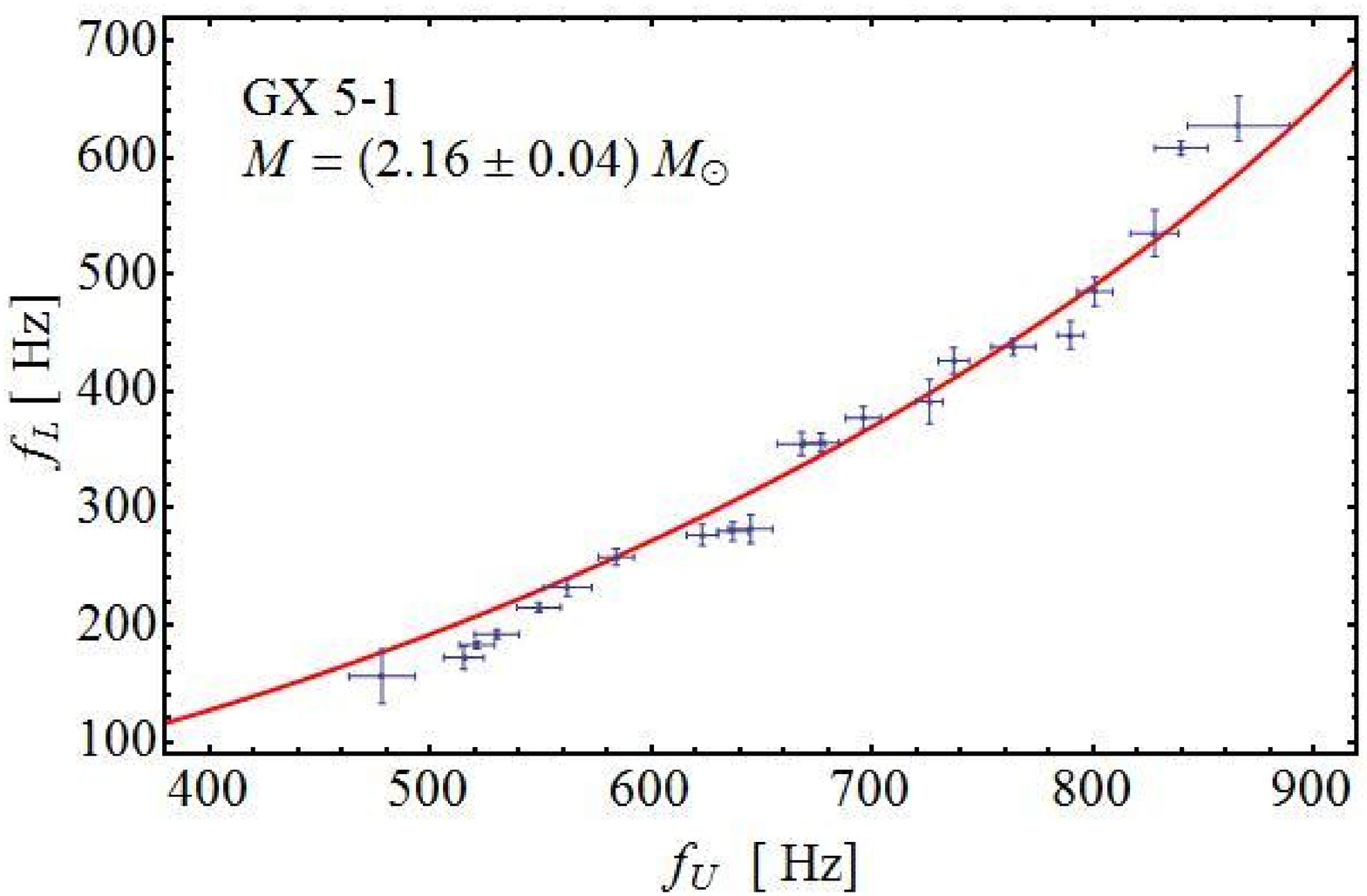}
		&
		\includegraphics[width=7 cm, height=6 cm, clip]{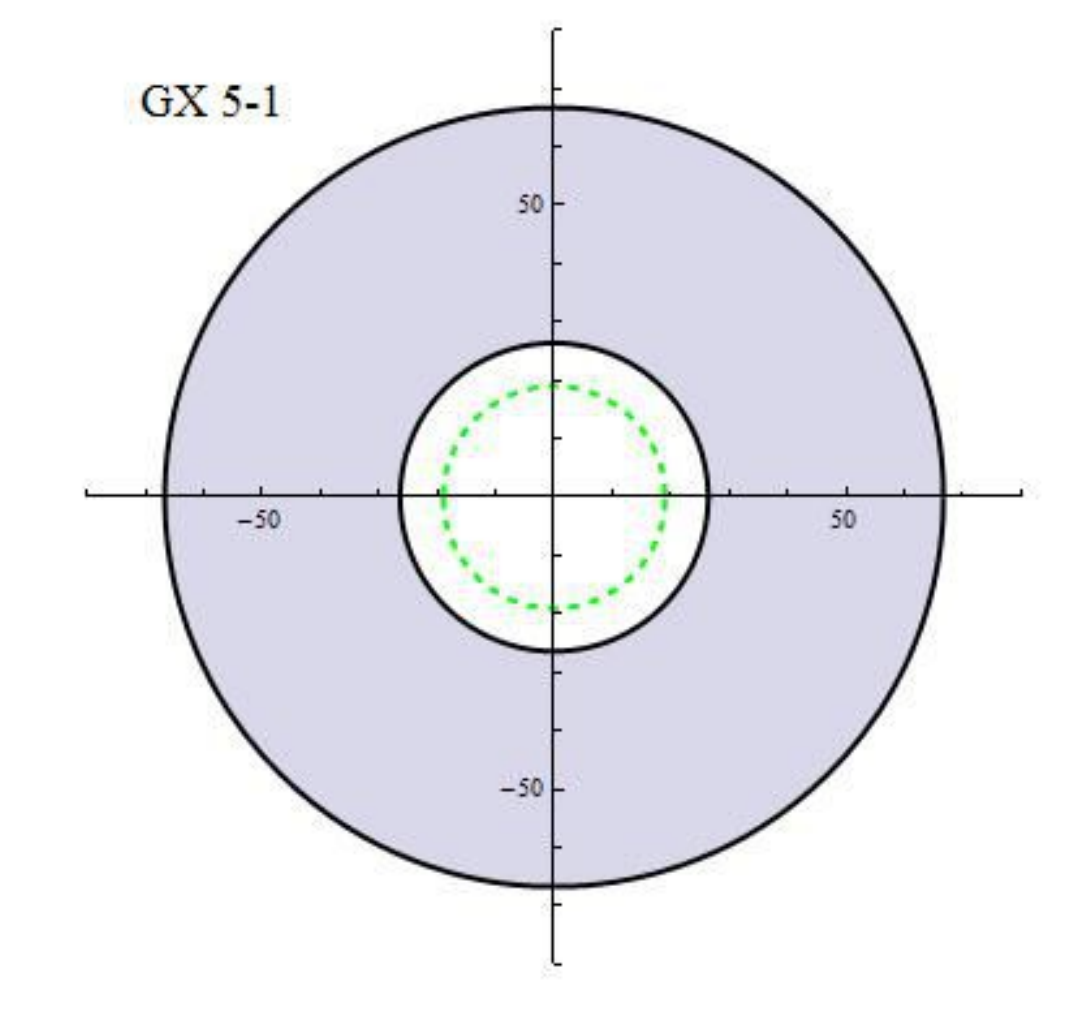}
	\end{tabular}
	\caption{Low mass X-ray binary system GX 5-1. The legend is the same as in Fig.1.}
	
\end{figure}

\begin{figure}[H]
	\centering
	\begin{tabular}{lr}
		\includegraphics[width=9 cm, height=6 cm, clip]{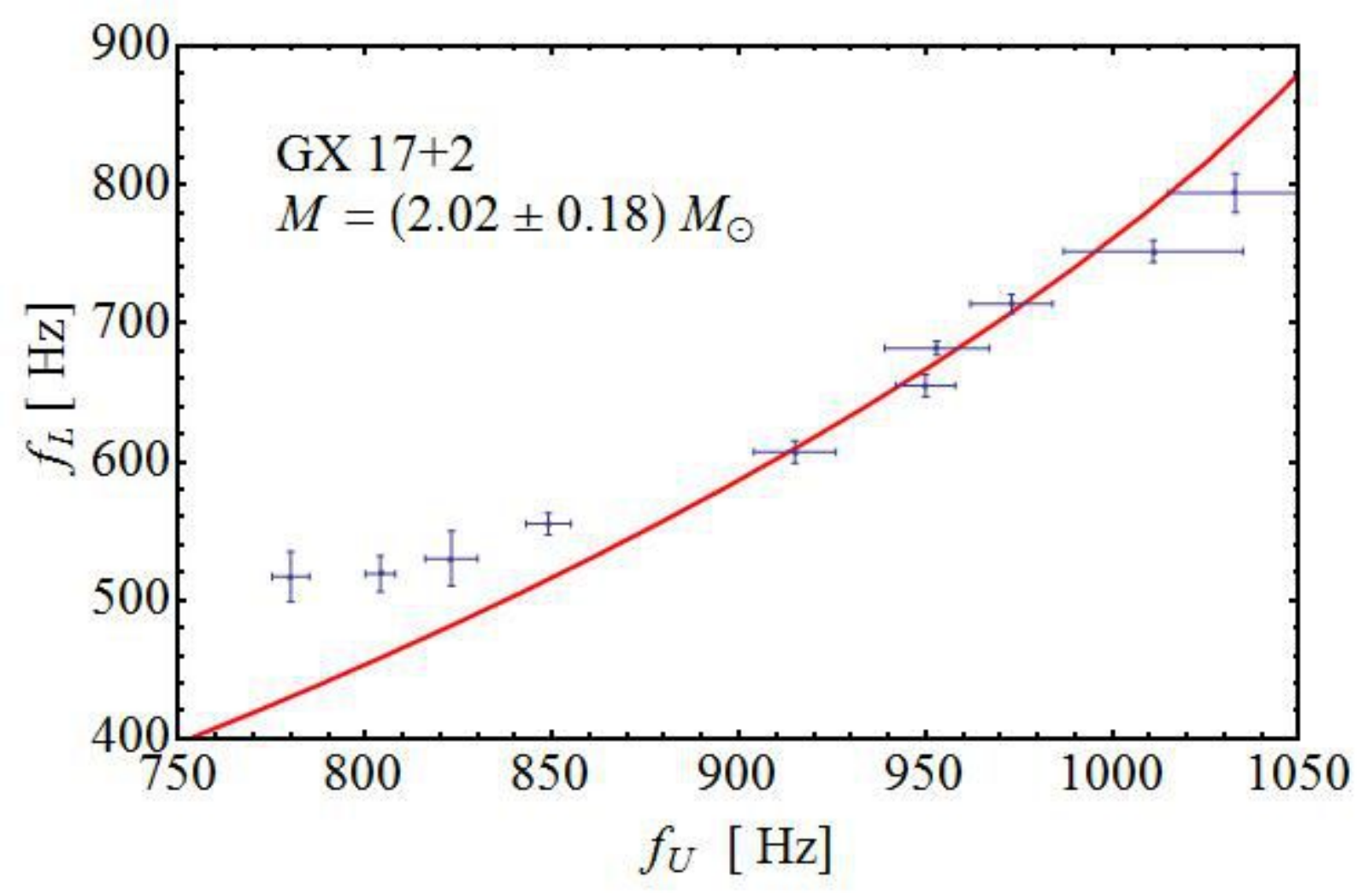}
		\includegraphics[width=6 cm, height=6 cm, clip]{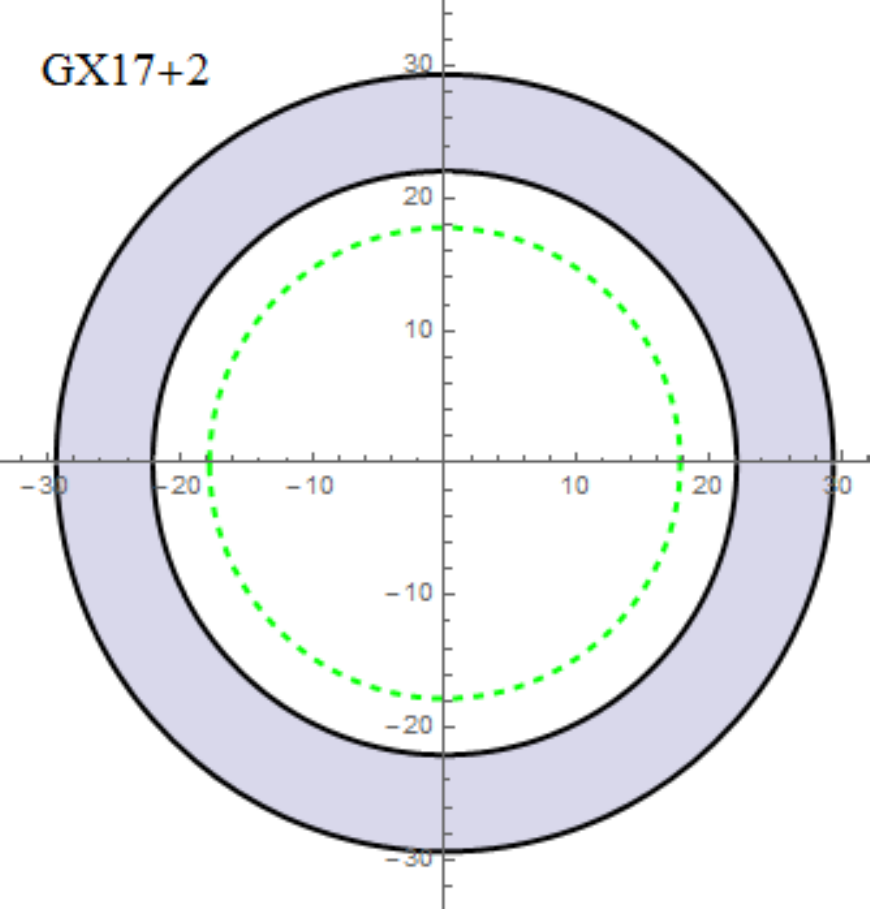}
	\end{tabular}
	\caption{Low mass X-ray binary system GX 17+2. The legend is the same as in Fig.1.}
	
\end{figure}

\begin{figure}[htbp]
	\centering
	\begin{tabular}{lr}
		\includegraphics[width=9 cm, height=6 cm, clip]{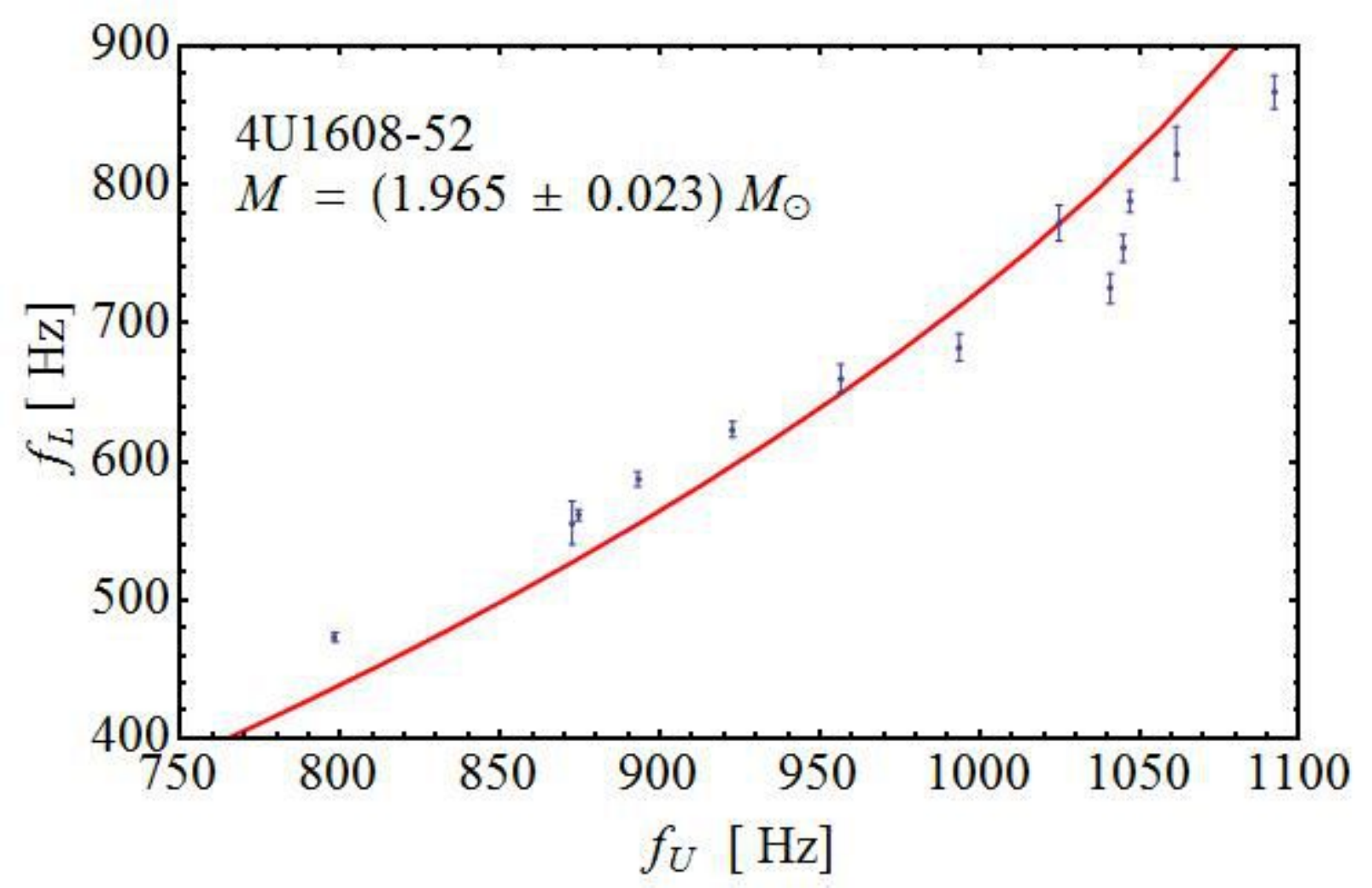}
		\includegraphics[width=7 cm, height=6 cm, clip]{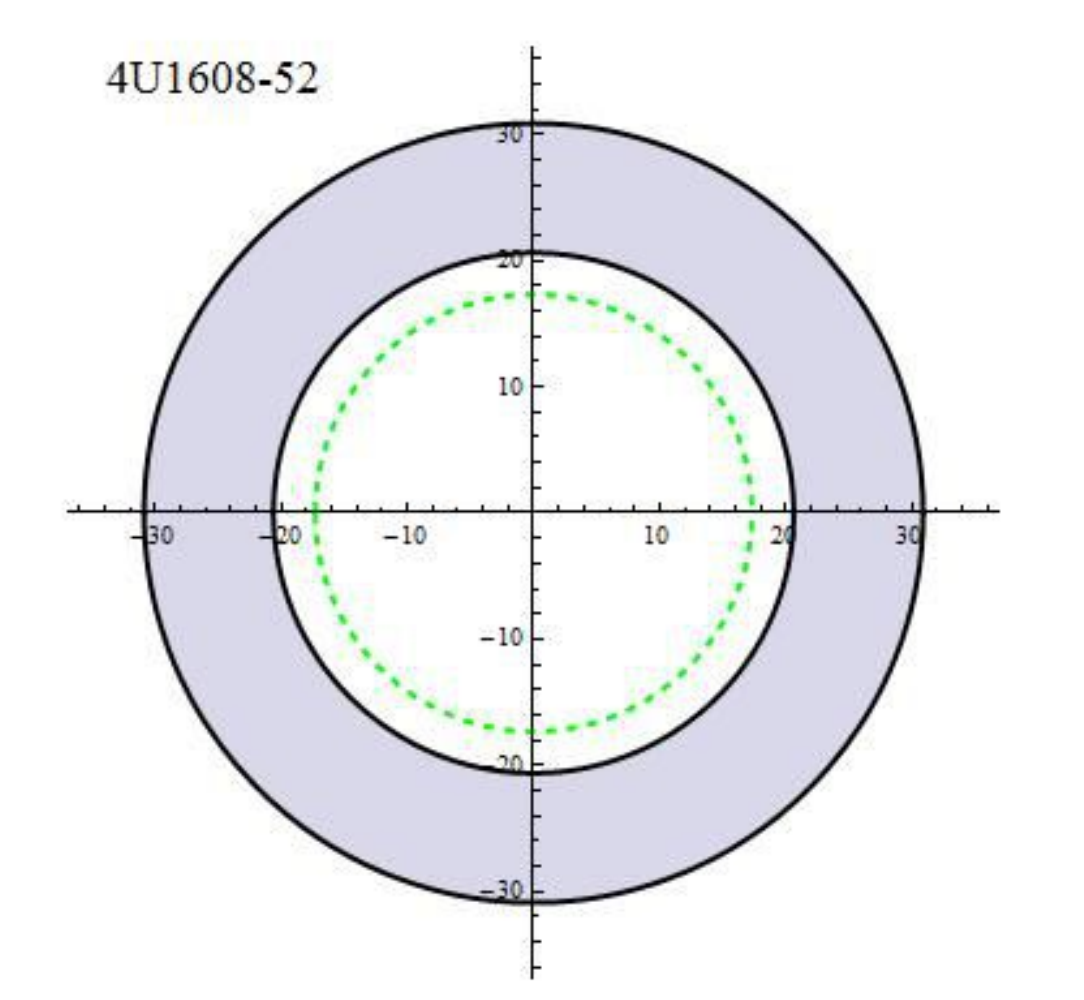}
	\end{tabular}
	\caption{Low mass X-ray binary system 4U1608-52. The legend is the same as in Fig.1. }
	
\end{figure}

\begin{figure}[H]
	\centering
	\begin{tabular}{cc}
		\includegraphics[width=9 cm, height=6 cm, clip]{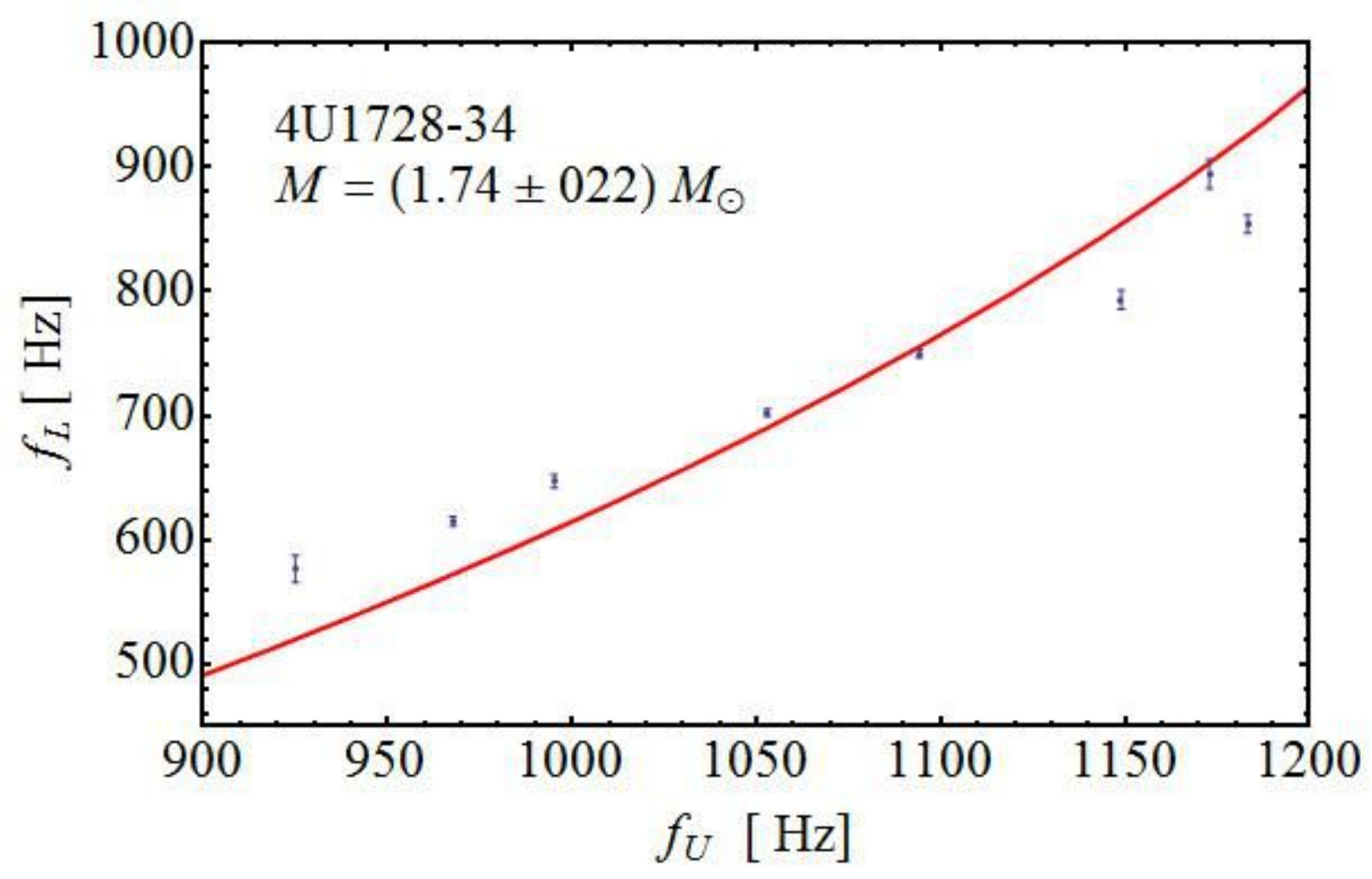}
		\includegraphics[width=7 cm, height=6 cm, clip]{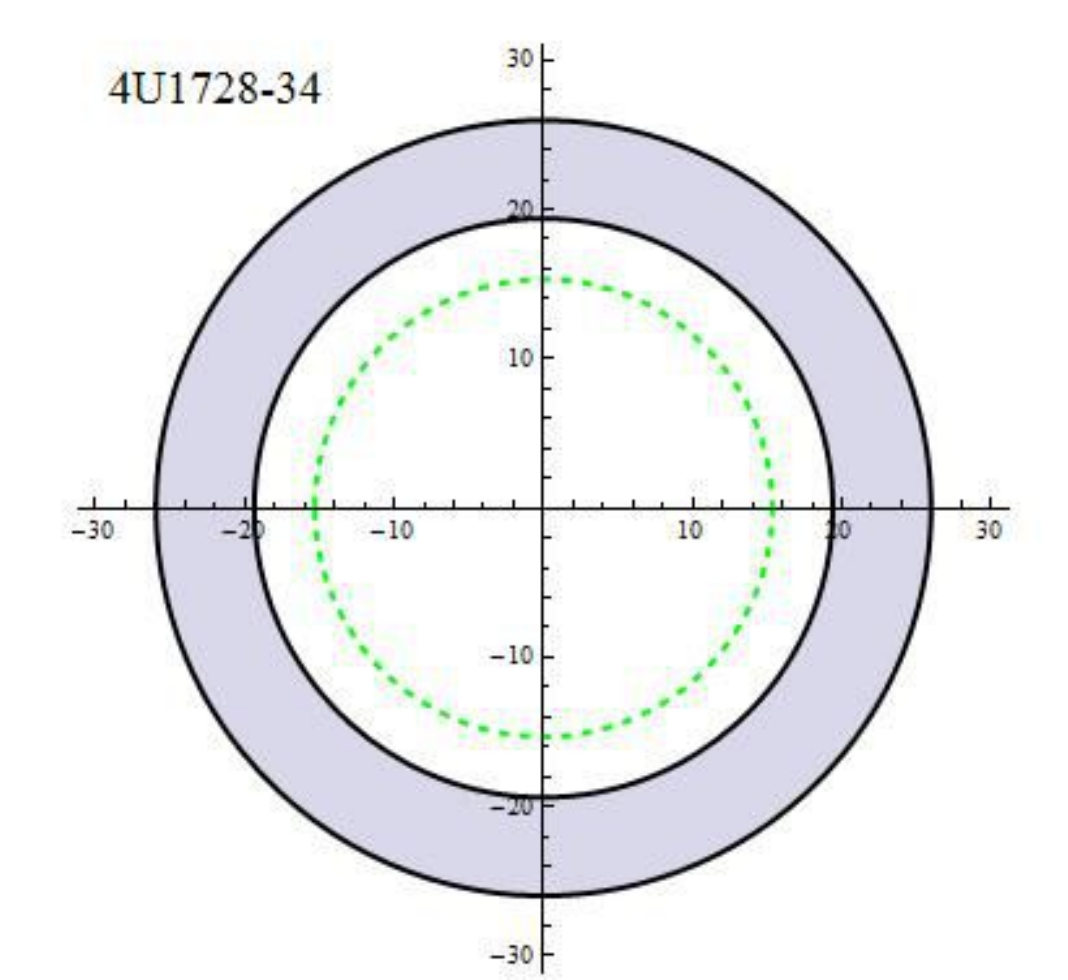}
	\end{tabular}
	\caption{Low mass X-ray binary system is 4U1728-34.The legend is the same as in Fig.1.}
	
\end{figure}

\begin{figure}[H]
	\centering
\includegraphics[width=14 cm, height=9 cm, clip]{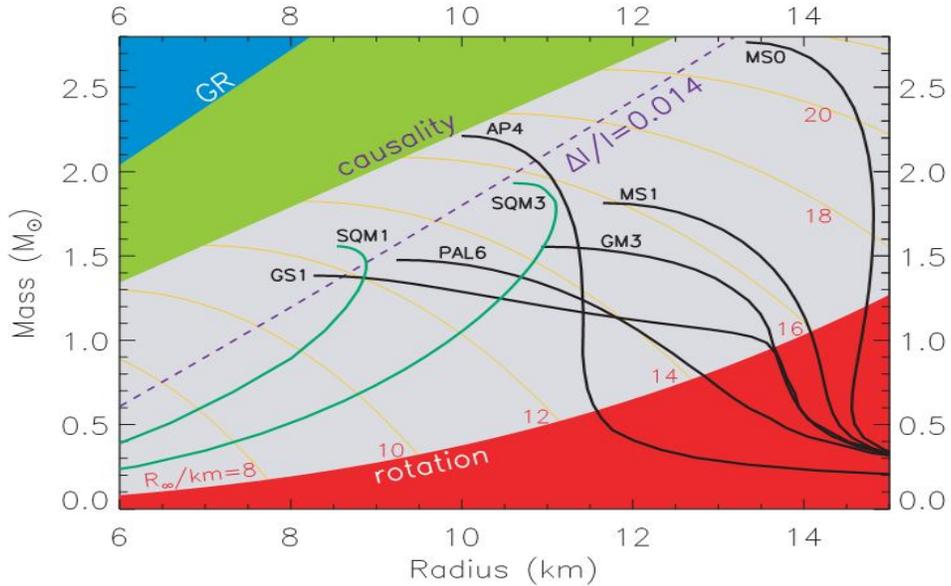}
\caption{The mass-radius relation for different equation of states for neutron stars \cite{28}.}	
\end{figure}

On left panel of Figs. 1-5 we have constructed the lower frequency versus the upper frequency using the fitting function and inferring the mass for Z (Cir X-1, GX 5-1, GX 17+2) and Atoll (4U1608-52, 4U1728-34) sources. In addition on right panel of Figs. 1-5 we show schematic representations of the structure of the accretion disks with their internal $R_{in}$ and external $R_{ex}$ radii (in kilometers). The width of the disks vary for all sources. One can show that the $R_{ISCO}$ is always less than $R_{in}$ and $R_{ex}$. The results are summarized in Table 1.

In Fig. 6 we show the mass-radius relations of Lattimer and Prakash \cite{28} constructed for static neutron stars with various equations of state. All stable configurations are inside the white region. If nuclear physics is correct, a measurement of $M$ and $R$ provides the internal composition of the NS i.e. one can rule out non-physical equations of state. Some masses are known with 99.9\% accuracy! Up to now observational constraints for the mass-radius relations of neutron stars favor stiff equations of state where one get masses larger than two solar mass. Additional observational data are needed to determine whether the equation of state is stiff or super-stiff \cite{29}.
In Table 1 we show the main results of this paper. Using formula (6) we have found radius of innermost stable circular orbits $r_{ISCO}$ for test particles in the gravitational field of compact objects. Internal $R_{in}$  and external $R_{ex}$ radii of the accretion disks have been inferred from using the lowest and highest values of the Keplerian frequencies from the QPO data. If one assumes that the central object in the LMXBs is a neutron star, then knowing mass $M$ of the sources one can show the range for the radius $R$ of neutron stars approximately using the Lattimer and Prakash diagram. If one assumes a black as a central compact object in the LMXBs, then using the Schwarzschild radius one can find the event horizon radius.

Table 1. Main characteristics of the compact objects and accretion disks

\begin{figure}[H]
	\centering
	\includegraphics[width=13 cm, height=5 cm, clip]{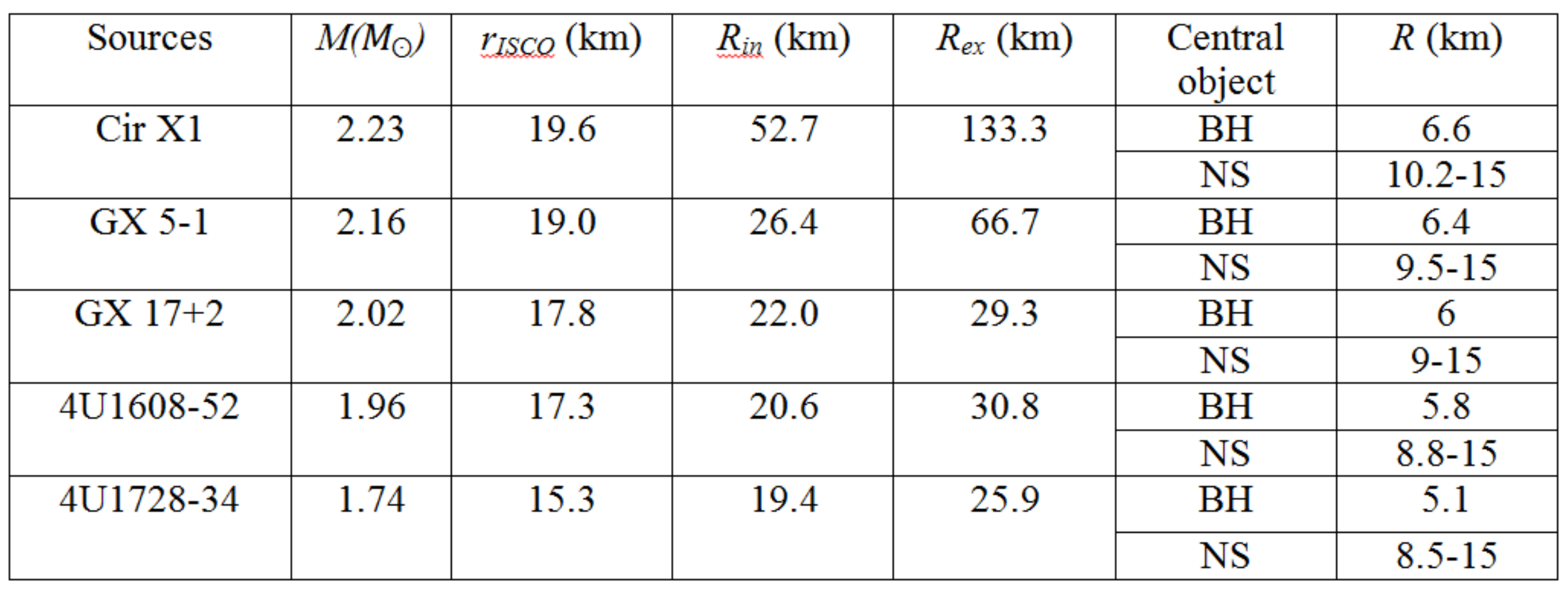}
\end{figure}
\section{Conclusion}

In this paper we study the quasi-periodic oscillation from low-mass X-ray binary systems. To explain the quasi-periodic oscillations in our work we exploit the relativistic precession model, because it is one of the simplest models, that involves the minimum set of free parameters.
In light of the relativistic precession model, we presented a detailed analysis. We considered the kilohertz quasi-periodic oscillations in the field of a static central object. Using analytic formulas for fundamental frequencies in the Schwarzschild spacetime we interpret the kilohertz quasi-periodic oscillations of low-mass X-ray binaries of the Atoll and Z sources.
Particularly we perform analyses for Z and Atoll sources Cir X1, GX 5-1, GX 17+2, 4U1608-52, 4U1728-34. We show that the quasi-periodic oscillations data can provide information on the parameters, namely, the mass and radius of compact objects in low-mass X-ray binaries.
Furthermore, we calculated the external and internal radii of the accretion disk and the innermost stable circular orbit radius. Unfortunately, from observations it is hard to obtain precise values of LMXB masses. Nevertheless, the approach used in this work can be considered as an alternative approach inferring the mass of the compact object.
One can state that the value for the radius of the central compact object is realistic, because it is less than the radius of the innermost stable circular orbits $r_{ISCO}$  and, in turn,  $r_{ISCO}$  itself is less than inner and outer radii of the accretion disk. We assumed that our compact object is a non-rotating body; it could be either a neutron star or a black hole, but not a white dwarf since its mass exceeds the Chandrasekhar mass limit. The mass values suggest these LMXBs harbor neutron stars but not black holes, unless the neutron star critical mass for gravitational collapse is very small. However, the observation of two solar mass neutron star in PSR J0348+0432 \cite{30} puts this value as a firm lower limit to the mass of black holes form from stellar gravitational collapse.
In order to have further confirmation of the nature of the compact object as a neutron star or a black hole one should consider a more general solution for the Einstein field equations taking into account the mass $M$, the angular momentum $J$ and the quadrupole moment $Q$. It is well-known that for a rotating black hole we have the relation between the angular momentum and quadrupole moment $Q=J^2/M$ and for a rotating neutron star we have always this relation $Q>J^2/M$. After inferring the values of $M$, $J$ and $Q$, and comparing the relationship for $Q$ and $J$ one can determine that the compact object is a neutron star or a black hole for sure. The first attempt in this direction is taken in Refs. \cite{31,32}. If a neutron star nature for these objects will be confirmed, this calculation will serve in addition to confirm or reject whether some of them are indeed more massive than PSR J0348+0432 as Table 1 suggests. However this issue is out of the scope of the current work. It will be considered in our future works.



\end{document}